\documentclass[a4paper]{JHEP3}
\pdfoutput=1
\usepackage[latin1]{inputenc}
\usepackage{bbm}
\usepackage{bm}
\usepackage{graphicx}
\usepackage{epsfig}
\usepackage{subfigure}
\usepackage{latexsym}
\usepackage{amsmath}
\usepackage{amsfonts}
\usepackage{amssymb}
\usepackage{color}
\let\normalcolor\relax

\def\tr{\mathrm{Tr}}

\definecolor{rossoCP3}{cmyk}{0,.88,.77,.40}

\newcommand{\beq}{\begin{equation}}
\newcommand{\eeq}{\end{equation}}

\newcommand{\be}{\begin{eqnarray}}
\newcommand{\ee}{\end{eqnarray}}

\title{\textcolor{rossoCP3} {Minimal Supersymmetric Technicolor}}
\preprint{\it CP$^3$-Origins: 2010-01}
\author{Matti Antola\footnote{matti.antola@helsinki.fi}\\
Department of Physics and Helsinki Institute of Physics, 
P.O.Box 64, FI-000140, University of Helsinki, Finland}
\author{Stefano Di Chiara\footnote{dichiara@cp3.sdu.dk}\\
{ CP}$^{ \bf 3}${-Origins}, 
Campusvej 55, DK-5230 Odense M, Denmark}
\author{Francesco Sannino\footnote{sannino@cp3.sdu.dk}\\
{ CP}$^{ \bf 3}${-Origins}, 
Campusvej 55, DK-5230 Odense M, Denmark}
\author{Kimmo Tuominen \footnote{kimmo.tuominen@jyu.fi} \\
Department of Physics, 
P.O.Box 35 (YFL), FI-40014 University of Jyv\"askyl\"a, Finland, 
        \\ and \\
Helsinki Institute of Physics, 
P.O.~Box 64, FI-00014 University of Helsinki, Finland}

\abstract
{We introduce novel extensions of the Standard Model featuring a supersymmetric technicolor sector. First we consider ${\cal N}=4 $ Super Yang-Mills  which breaks to ${\cal N}=1$ via the electroweak (EW) interactions and coupling to the MSSM. This is a well defined, economical and calculable extension of the SM involving the smallest number of fields. It constitutes an explicit example of a natural supersymmetric conformal extension of the Standard Model featuring a well defined connection to string theory. It allows to interpolate, depending on how we break the underlying supersymmetry, between unparticle physics and Minimal Walking Technicolor. As a second alternative we consider other ${\cal N} =1$ extensions of the Minimal Walking Technicolor model. The new models allow all the standard model matter fields to acquire a mass.}
\keywords{Bosonic Technicolor, Supersymmetric Technicolor, $\mathcal{N}=4$ SUSY}

\begin{document}
\section{Minimal Walking Technicolor and Supersymmetry}
\label{introduction}

The Standard Model (SM) of particle interactions passes a large number of experimental tests. Yet we know that it cannot be the ultimate model of nature since it fails to explain the origin of matter-antimatter asymmetry and the abundance of cold dark matter. Several extensions of the SM have been proposed, and two stand out in the quest of a better theory: Supersymmetry and Technicolor. 

The appeal of Supersymmetry (SUSY) resides in its higher level of space-time symmetries as well as in its often praised natural link to string theory. The most investigated route to introduce SUSY has been to supersymmetrize the SM and then invoke some mechanism to break SUSY again, given that no sign of the superpartners has yet been observed in experiments. 

Technicolor declares the Higgs sector of the SM to be a low energy effective theory in which the Higgs is not elementary but composite. Electroweak symmetry breaking is triggered by a dynamical low energy condensate. The mechanism is automatically insensitive to high energy physics\footnote{See \cite{Sannino:2009za,Sannino:2008ha} for up-to-date reviews and \cite{Hill:2002ap} for a review of the older models.}. The main appeal of technicolor is that we have already encountered similar phenomena in nature: superconductivity is a time honored example while relativistic version is the spontaneous chiral symmetry breaking in the vacuum of the ordinary Quantum Chromodynamics (QCD). The earliest models of technicolor were found to have problems with the EW precision data. However, recent developments led to models that have been shown to pass the precision tests \cite{Sannino:2004qp,Dietrich:2005jn}.

Technicolor predicts the existence of a tower of massive states whose mass is of the order of the EW scale, although pseudo-Goldstone bosons can be lighter. This fact naturally explains why we have not detected technicolor yet. To give masses to the SM fermions one must, however, resort to another unknown sector, usually defined only in the low energy limit by an effective theory.  In this work we consider a supersymmetric theory. One motivation for this is that we can use fundamental scalars to either transmit spontaneous symmetry breaking to the SM fermion sector, or even to drive electroweak symmetry breaking. 

The supersymmetric technicolor idea was put forward in \cite{Dine:1981za}, though the phenomenological viability of early models seemed difficult to achieve. An important difference with our models is that the underlying supersymmetric and technicolor theories, which can be resumed by decoupling either the technicolor fields or the superpartners, are both phenomenologically viable\footnote{However, in case we give up SUSY, we should introduce an alternative extended technicolor sector to generate the SM fermion masses.}.  These underlying theories are, respectively, the minimal supersymmetric standard model (MSSM) and the minimal walking technicolor (MWT) model \cite{Sannino:2004qp,Dietrich:2005jn,Evans:2005pu}. Also, we are not aiming at breaking supersymmetry dynamically.

In MWT \cite{Sannino:2008ha} the gauge group is $SU(2)_{TC}\times SU(3)_C\times SU(2)_L\times U(1)_Y$ and the field content of the technicolor sector is constituted by four techni-fermions and one techni-gluon all in the adjoint representation of $SU(2)_{TC}$.  The model features also a pair of Dirac leptons, whose left-handed components are assembled in a weak doublet, necessary to cancel the Witten anomaly \cite{Witten:1982fp} arising when gauging the new technifermions with respect to the weak interactions. Summarizing, the fermionic particle content of the MWT is given explicitly by
\beq Q_L^a=\left(\begin{array}{c} U^{a} \\D^{a} \end{array}\right)_L , \qquad U_R^a \
, \quad D_R^a, \quad a=1,2,3 \ ; \quad
L_L=\left(
\begin{array}{c} N \\ E \end{array} \right)_L , \qquad N_R \ ,~E_R \ . 
\label{MWTp}
\eeq 
The following generic hypercharge assignment is free from gauge anomalies:
\begin{align}
Y(Q_L)=&\frac{y}{2} \ ,&\qquad Y(U_R,D_R)&=\left(\frac{y+1}{2},\frac{y-1}{2}\right) \ ,  \nonumber\\
Y(L_L)=& -3\frac{y}{2} \ ,&\qquad
Y(N_R,E_R)&=\left(\frac{-3y+1}{2},\frac{-3y-1}{2}\right) \ \label{assign2} .
\end{align}
The global symmetry of this technicolor theory is $SU(4)$, which breaks explicitly to $SU(2)_L \times U(1)_Y$ by the natural choice of the EW embedding \cite{Sannino:2004qp,Dietrich:2005jn}. EWSB is triggered by a fermion bilinear condensate and the vacuum choice is stable against the SM quantum corrections \cite{Dietrich:2009ix}.

The supersymmetric version of MWT depends on the hypercharge parameter $y$. For $y=\pm 1$, one can construct an approximately ${\cal N}=4$ supersymmetric technicolor sector, which is broken down to ${ \cal N}=1$ SUSY only by EW gauge and Yukawa interaction terms. We build the model in section \ref{n4model} and name it {\it Minimal Supersymmetric Conformal Technicolor} (MSCT). We study it briefly in two different regimes to understand the main constraints and challenges.

In section \ref{n1model} we consider other  ${\cal N} =1 $ supersymmetric extensions of MWT and two specific choices of the hypercharge of the technifermion matter. We call this extension {\it Minimal Supersymmetric Technicolor} (MST). This extension is less economical than the MSCT in the number of fields needed but has other interesting properties such as the presence of a gauge singlet superfield, which can be used to solve the $\mu$ problem of the MSSM, and a Higgs scalar candidate already within the spectrum of MWT superpartners. 

The basic properties of the models we are about to introduce are: 
\begin{itemize}
\item The models are supersymmetric. Thus low energy physics is not sensitive to high energy physics and we call the models {\it natural}.
\item Supersymmetry is broken softly.
\item The particle contents are given by that of the MSSM (including its two Higgs fields) plus the supersymmetric technicolor sector.
\item The models can interpolate between different extensions of the SM at the TeV scale, such as unparticle \cite{Georgi:2007ek,Georgi:2007si}, MWT \cite{Sannino:2004qp,Dietrich:2005jn}, and MSSM (see \cite{Martin:1997ns} for a review).
\item The models are complete theories in which electroweak symmetry breaking (EWSB) is triggered either by a dynamical fermion condensate or by the non-zero vacuum expectation value (vev) of an elementary scalar. Electroweak symmetry breaking is transmitted to the SM fermion sector via elementary scalars.
\item The models can possibly solve the $\mu$ problem, or the flavor-changing and CP-breaking problems of the SUSY breaking sector within MSSM.
\item The MSCT model possesses an approximate $\mathcal{N}=4$ supersymmetry, and can be viewed as a $\mathcal{N}=4$ sector coupled to the MSSM.
\item The MSCT model possesses a clear and direct link to string theory in such a way that AdS/CFT techniques \cite{Maldacena:1997re} are readily applicable to realistic extensions of the SM.
\end{itemize}

Other possibilities of giving masses to standard model fermions within MWT have been considered in the literature earlier. An explicit construction of an extended technicolor type model appeared in \cite{Evans:2005pu}. A less natural model introducing a scalar ({\it bosonic technicolor}) mimicking the effects of the extended technicolor interactions has also been introduced in \cite{Antola:2009wq,Zerwekh:2009yu} following the pioneering work of Simmons  \cite{Simmons:1988fu},  Kagan and Samuel \cite{Kagan:1991gh}, and Carone  \cite{Carone:1992rh,Carone:1994mx}. More recently this type of models have been investigated also in \cite{Hemmige:2001vq,Carone:2006wj,Zerwekh:2009yu}. Interesting related work can be also found in \cite{Chivukula:1990bc,Samuel:1990dq,Dine:1990jd,Kagan:1990az,Kagan:1991ng,Dobrescu:1995gz,Dobrescu:1998ci,Chivukula:2009ck}. 

\section{Minimal Supersymmetric Conformal Technicolor (MSCT)}
\label{n4model}

To build the supersymmetric technicolor theory we must supersymmetrize the additional technicolor sector, given by MWT, and also the standard model. We start by noting that the fermionic and gluonic spectrum of MWT fits perfectly in an ${\cal N}=4$ supermultiplet.
In fact the $SU(4)$ global symmetry of MWT is nothing but the well known $SU(4)_R$ $R$ symmetry of the ${\cal N}=4$ Super Yang-Mills (4SYM) theory. Having at hand already a great deal of the spectrum of 4SYM we explore the possibility of using this theory as a natural candidate for supersymmetric technicolor. For the reader's convenience we have summarized the 4SYM Lagrangian in terms of the ${\cal N}=1$ superfields, and in elementary field components in Appendix \ref{N4susyl}. We refer to this appendix for the basic properties of the 4SYM theory, Lagrangian and notation.

We gauge part of the $SU(4)_R$ global symmetry of the supersymmetric technicolor theory in order to couple the new supersymmetric sector to the weak and hypercharge interactions of the SM. We choose to do this in such a way that the model can still preserve ${\cal N} =1 $ SUSY. To this end one of the four Weyl technifermions $\bar{U}_R$, $\bar{D}_R$, $U_L$, $D_L$ is identified with the techni-gaugino and should be a singlet under the SM gauge group. The only possible candidates for this role are $\bar{U}_R$ and $\bar{D}_R$, for $y=\mp 1$ respectively: we arbitrarily choose $y=1$ and identify $\bar{D}_R$ with the techni-gaugino. With this choice the charge assignments of the technicermions ($Q_L$ is the left-handed doublet of $U_L$ and $D_L$) and new leptons ($L_L$ is the left-handed doublet of $N_L$ and $E_L$) under $SU(2)_{TC}\times SU(3)_C\times SU(2)_L\times U(1)_Y$ are
\be
&&Q_L\sim \left(\textbf{3},\textbf{1},\textbf{2},\frac{1}{2}\right),\ \bar{U}_R\sim(\textbf{3},\textbf{1},\textbf{1},-1),\ \bar{D}_R\sim(\textbf{3},\textbf{1},\textbf{1},0),\ \nonumber\\
&&L_L\sim\left(\textbf{1},\textbf{1},\textbf{2},-\frac{3}{2}\right),\ \bar{N}_R\sim(\textbf{1},\textbf{1},\textbf{1},1),\ \bar{E}_R\sim(\textbf{1},\textbf{1},\textbf{1},2). 
\label{chaa}
\ee
Based on these assignments we then define the scalar and fermion components of the ${\cal N}=4$ superfields via 
\beq
\left(\tilde{U}_L,\ U_L\right)\in \Phi_1,\quad \left(\tilde{D}_L,\ D_L\right)\in \Phi_2,\quad   
\left(\tilde{\bar{U}}_R,\ \bar{U}_R\right)\in \Phi_3,\quad \left(G,\ \bar{D}_R\right)\in V,  
\label{superq}
\eeq
where we used a tilde to label the scalar superpartner of each fermion. We indicated with $\Phi_i$, $i=1,2,3$ the three chiral superfields of 4SYM and with $V$ the vector superfield.  Four more chiral superfields are necessary to fully supersymmetrize the MWT model, i.e.:
\beq
\left(\tilde{N}_L,\ N_L\right)\in \Lambda_1,\quad \left(\tilde{E}_L,\ E_L\right)\in \Lambda_2,\quad   
\left(\tilde{\bar{N}}_R,\ \bar{N}_R\right)\in N,\quad\left(\tilde{\bar{E}}_R,\ \bar{E}_R\right)\in E.
\label{superl}
\eeq

As one can see from the spectrum in Eq.(\ref{chaa})  there is no scalar field that can be coupled to SM fermions in a gauge invariant way and play the role of the SM Higgs boson (a weak doublet with hypercharge $Y=\pm\frac{1}{2}$). We therefore introduce in the theory two Higgs doublet superfields with respective charge assignment
\beq
H\sim \left(\textbf{1},\textbf{1},\textbf{2},\frac{1}{2}\right),\ H^{\prime}\sim\left(\textbf{1},\textbf{1},\textbf{2},-\frac{1}{2}\right),
\label{superH}
\eeq
where the presence of both $Y=\pm\frac{1}{2}$ superfields is needed to give mass by gauge invariant Yukawa terms to both the upper and lower components of the weak doublets of SM fermions. With this choice it is rather natural to take the MSSM to describe the supersymmetric extension of the Higgsless SM sector.

Thus the model's full particle content is given by that of the MSSM and the new technicolor fields and their superpartners. All the MSSM fields are defined as singlets under $SU(2)_{TC}$. The resulting MSCT model is naturally anomaly-free, since both the MWT and the MSSM are such. We summarize in Table \ref{MSCTsuperfields} the quantum numbers of the superfields in Eqs.(\ref{superq},\ref{superl},\ref{superH}).
\begin{table}[h]
\centering
\begin{tabular}{c|c|c|c|c}
\noalign{\vskip\doublerulesep}
Superfield & SU$(2)_{TC}$ & SU$(3)_{\text{c}}$ & SU$(2)_{\text{L}}$ & U$(1)_{\text{Y}}$\tabularnewline[\doublerulesep]
\hline
\noalign{\vskip\doublerulesep}
$\Phi_{1,2}$ & Adj & $1$ & $\square$ & 1/2\tabularnewline[\doublerulesep]
\noalign{\vskip\doublerulesep}
$\Phi_{3}$ & Adj & 1 & 1 & -1\tabularnewline[\doublerulesep]
\noalign{\vskip\doublerulesep}
$V$ & Adj & 1 & 1 & 0\tabularnewline[\doublerulesep]
\noalign{\vskip\doublerulesep}
$\Lambda_{1,2}$ & 1 & 1 & $\square$ & -3/2\tabularnewline[\doublerulesep]
\noalign{\vskip\doublerulesep}
$N$ & 1 & 1 & 1 & 1\tabularnewline[\doublerulesep]
\noalign{\vskip\doublerulesep}
$E$ & 1 & 1 & 1 & 2\tabularnewline[\doublerulesep]
\noalign{\vskip\doublerulesep}
$H$ & 1 & 1 & $\square$ & 1/2\tabularnewline[\doublerulesep]
\noalign{\vskip\doublerulesep}
$H'$ & 1 & 1 & $\square$ & -1/2\tabularnewline[\doublerulesep]
\end{tabular}\caption{MSCT ${\cal N}=1$ superfields}
\label{MSCTsuperfields}
\end{table}

The renormalizable superpotential for the MSCT, allowed by gauge invariance, and which we require additionally to be ${\cal N}=4$ invariant in the limit of $g_{TC}$ much greater than the other coupling constants and to conserve baryon and lepton numbers\footnote{We assume all the MWT particles to have baryon and lepton numbers equal to zero.}, is
\beq
P=P_{MSSM}+P_{TC},
\label{spot}
\eeq
where $P_{MSSM}$ is the MSSM superpotential, and
\beq
P_{TC}=-\frac{g_{TC}}{3\sqrt{2}} \epsilon_{ijk} \epsilon^{abc} \Phi^a_i \Phi^b_j \Phi^c_k+y_U \epsilon_{ij3}\Phi^a_i H_j\Phi^a_3+y_N \epsilon_{ij3}\Lambda_i H_j N+y_E \epsilon_{ij3}\Lambda_i H^{\prime}_j E+y_R \Phi^a_3 \Phi^a_3 E.
\label{spmwt}
\eeq
Relaxing the requirement of $\mathcal{N}=4$ invariance, the coefficient of the first term would be a general Yukawa $y_{TC}$. We have investigated the running of such a coupling and we found that it tends towards $g_{TC}$ at low energies \cite{Antola:2010jk}, which also confirms the result of \cite{Petrini:1997kk}. This justifies our choice to set it equal to the technicolor gauge coupling itself.

One of the interesting general features of the MSCT spectrum is the existence of a supersymmetric fourth family of leptons. The MWT predicts the natural occurrence of a fourth family of leptons around the EW energy scale, put forward first in \cite{Dietrich:2005jn}. The physics of these fourth family of leptons has been studied in \cite{Antipin:2009ks,Frandsen:2009fs}. From the EW  point of view there is little difference between the MWT and a fourth-family extended SM at the EW scale. Since the MSCT is a supersymmetrized version of the MWT the former now features, besides the techniquarks, a natural super 4th family of leptons awaiting to be discovered at colliders, albeit with more exotic electric charges: the new electron will be doubly charged \cite{Antipin:2010it}.

\subsection{The MSCT Landscape}

Many features of the spectrum depend on which part of the MSCT parameter space one chooses to study. The MSCT allows model builders to investigate a number of inequivalent extensions of the (MS)SM. These inequivalent extensions are determined by the choice of the value of the coupling constant $g_{TC}$ of the supersymmetric technicolor sector near the EW scale as well as by the vacuum choice permitted by the flat directions and by the SUSY breaking pattern.

It is not possible to exhaust in this work all the possibilities and, hence, we limit ourselves here to introduce the idea and the basic features. We identify two basic regimes, which both have to be augmented with soft SUSY breaking: First, the perturbative one, in which the supersymmetric technicolor coupling $g_{TC}$ is sufficiently small allowing  the new sector to be treated in perturbation theory and denote this model with pMSCT. Second, we discuss the case in which the supersymmetric technicolor is strongly coupled and we will denote it as sMSCT. Each specific model deserves to be studied on its own and some of these models will be investigated in  more detail in future publications.

A third regime, which we will not discuss further in this paper, is if SUSY is not broken and the supersymmetric technicolor dynamics is strongly coupled at the EW scale. Then we must use non-perturbative methods to investigate the effects of the new sector on the MSSM dynamics and vice versa. For example, we can no longer  use the single particle state interpretation in terms of the underlying degrees of freedom of the supersymmetric technicolor model but rather must use an {\it unparticle} language given that the supersymmetric technicolor model is exactly conformal, before coupling it to the MSSM.  The model resembles the one proposed in \cite{Sannino:2008nv} in which, besides a technicolor sector, one has also coupled a natural composite unparticle composite. If no SUSY breaking terms are added directly to the 4SYM sector then conformality will be broken only via weak and hypercharge interactions. An important further point is that one can use the machinery of the AdS/CFT correspondence to make reliable computations in the nonperturbative sector, considering the effects of the EW interactions as small perturbations. 

\subsection{Perturbative MSCT (pMSCT)}

The simplest case to consider is the one in which the new sector is weakly coupled at the EW scale and can be treated perturbatively. In this case the spectrum of states, which can be observed at the EW scale, is constituted by the elementary fields introduced in \eqref{superq} and \eqref{superl}, plus the MSSM ones. The detailed mass spectrum depends on the structure of the SUSY breaking terms and on the corrections induced by the EW symmetry on the supersymmetric technicolor sector.

The spectrum is rich with several novel weakly coupled particles, such as the new techni-up and techni-down, and their respective superpartners, which can emerge at the LHC. The superpartners will be similar to ordinary squarks but will carry technicolor instead of color. The weak processes involving the production of squarks at colliders should be re-investigated to take into account the presence of these new states; in this paper we carry out a basic analysis of the masses of fermion states.

We note that assigning a nonzero vev only to the Higgs scalars does not generate a mass for the neutralino $D_L$. Since it couples to the $Z$, a massless $D_L$ would have been already observed at LEP. To fix this, one can break $SU(2)_{TC}\times SU(2)_L\times U(1)_Y$ down to $U(1)_{TC}\times U(1)_{EM}$ by assigning a nonzero vev to the {\it techni-Higgs} $\tilde{D}_L$. Then, due to the Yukawa coupling generated by the first term in the superpotential, Eq.(\ref{spmwt}), the neutralino $D_L$ becomes massive after symmetry breaking.  Thus electroweak symmetry breaking is driven by the negative mass squared of $\tilde{D}_L$, $\tilde{H}$ and $\tilde{H}'$\footnote{Notice that we indicate the scalar component of each weak doublet superfield with a tilde.}.

With the above vacuum condensates the electroweak gauge group breaks down to electromagnetism and simultaneously the technicolor gauge group down to $U(1)_{TC}$. The phenomenological constraints on a new  $U(1)$ massless gauge boson were studied in \cite{Dobrescu:2004wz}. The lower limits on the scale of dimension six operators with SM fields and the new massless photon are in the TeV range.

We choose the vacuum expectation value (vev) of the techniscalar to be aligned in the third direction of the $SU(2)_{TC}$ gauge space. We define the vevs:
\beq
\left<\tilde{D}_L^3\right>=\frac{v_{TC}}{\sqrt{2}},\quad  \left<\tilde{H}_0\right>=s_\beta \frac{v_H}{\sqrt{2}},\quad  \left<\tilde{H}_0^\prime\right>=c_\beta \frac{v_H}{\sqrt{2}},
\label{veveq}
\eeq
where all the vevs are chosen to be real, $s_\beta=\sin\beta$, and $c_\beta=\cos\beta$.

After EWSB the mass terms of gauge bosons are written as a function of the mass eigenstates as:
\beq
-{\cal L}_{g\textrm{-}mass}=g_{TC}^2 v_{TC}^2 G^{+}_\mu G^{-\mu}+\frac{g^2_L}{2} \left( v_{TC}^2+ v_H^2\right) W_\mu^+ W^{-\mu}+\frac{g^2_L+g^2_Y}{4} \left( v_{TC}^2+ v_H^2\right) Z_\mu Z^\mu
\label{gmL}
\eeq
where
\beq
G^{\pm}_\mu=\frac{1}{\sqrt{2}}\left( G^1_\mu \mp i\, G^2_\mu \right)\ ,\ W^{\pm}_\mu=\frac{1}{\sqrt{2}}\left( W^1_\mu \mp i\, W^2_\mu \right)\ ,\  Z_\mu=c_w W^3_\mu-s_w B\ ,\ t_w=\frac{g_Y}{g_L}.
\label{gme}
\eeq
The $\pm$ exponent of the techni-gluon refers to the $U(1)_{TC}$ charge, while the $\pm$ exponent on the EW gauge bosons refer to the usual EM charge. We see the electroweak scale is set by 

\beq
\sqrt{v^2_{H}+v_{TC}^2}=246 \textrm{ GeV}.
\eeq

The remaining, massless states are the techni-photon and the EW photon:
\beq
G_\mu=G^3_\mu\ ,\qquad  A_\mu=s_w W^3_\mu + c_w B\
\label{g0e}
\eeq

Focusing on the fermion spectrum, the lower bounds on the mass of the lightest neutralino and chargino are \cite{Amsler:2008zzb}:
\beq
\label{exmass}
m_{\chi_0^0}>46 \textrm{ GeV}\,,\ m_{\chi_0^\pm}>94 \textrm{ GeV}\,. 
\eeq

These limits refer to the MSSM, but are rather general, since they are extracted mostly from the $Z$ decay to neutralino-antineutralino pair for the former, and from photo-production of a chargino-antichargino pair at LEPII for the latter. We can therefore assume these limits to hold also for the MSCT. Because of their generality and independence from the coupling strength (as long as it is not negligible), we use the lower bound on the chargino mass also for the mass of the doubly-charged chargino $E$. Note that the presence of the term proportional to $y_R$ in the superpotential, Eq.(\ref{spmwt}), allows it to decay into singly charged ordinary particles, thereby escaping cosmological constraints on charged stable particles.

The techni-gaugino $\bar{D}^3_R$ is an EW singlet fermion with zero charge under $U(1)_{TC}$ and therefore plays the role of a sterile right-handed neutrino, which can be very light. Because the mass of the lightest techineutralino is a decreasing function of the mass of $\bar{D}^3_R$, we can assume the soft SUSY breaking mass $M_D$ of the techni-gaugino to be small. This assumption turns out to be phenomenologically favored.

\begin{figure}[h]
\begin{center}
\includegraphics[width=18pc]{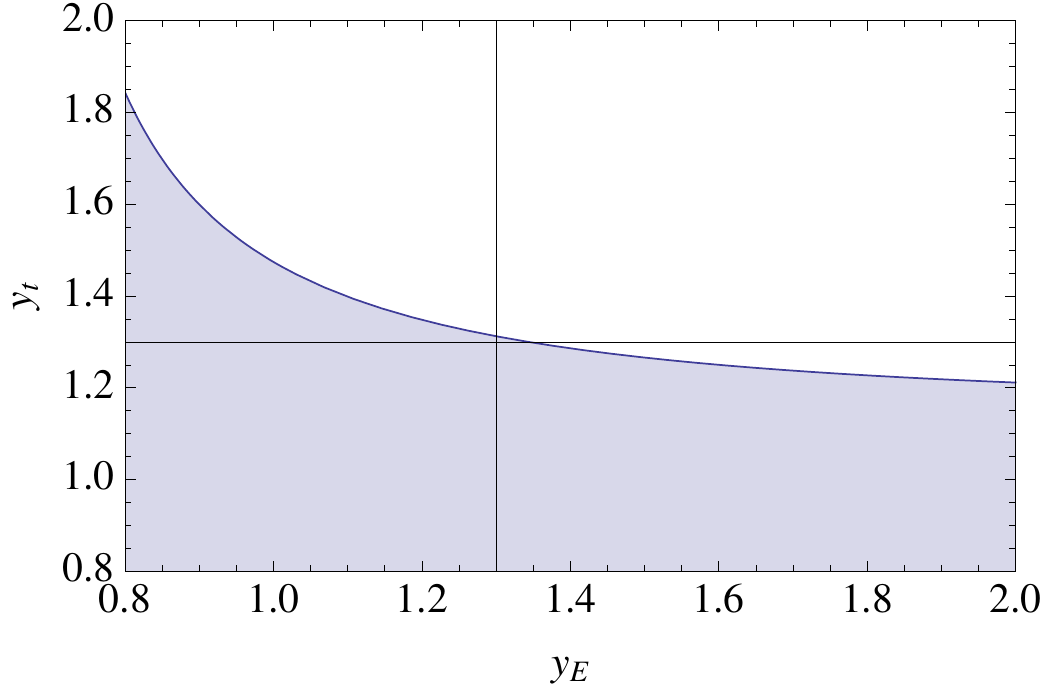}
\caption{\label{fig:ytyE}Shaded area shows experimentally excluded values of the Yukawa couplings $y_t$ and $y_E$ in pMSCT model.}
\vspace{1pc}
\end{center}
\end{figure}

Other useful limits on the parameters are obtained by using the fact that the smallest eigenvalue of a semi-positive definite square matrix is smaller or equal to any eigenvalue of the principal submatrices. By using the neutralino, chargino, and doubly charged chargino mass matrices, we get for example

\beq
\label{netb}
v_{TC}>2 \frac{46\textrm{ GeV}}{\sqrt{g_L^2+g_Y^2}}=124\textrm{ GeV}\,,\ v_H<213\textrm{ GeV}\,,\ m_{cc}=\frac{y_E c_{\beta } v_H}{\sqrt{2}}>94 \textrm{ GeV}\,,\ m_t=\frac{y_t}{y_E}t_\beta m_{cc}
\eeq
where the subscript $t$  refers to the top quark and $m_{cc}$ is the doubly charged lepton mass. One of the most important inequalities is
\beq
\label{ytyE}
y_t > \frac{173}{213} \sqrt{\frac{1}{\frac{1}{2}-\frac{94^2}{y_E^2 213^2}}}\,.
\eeq
This last bound is plotted in Figure \ref{fig:ytyE}, where the shaded area shows the values of $y_t$ and $y_E$ excluded by the experiment: it is evident that either $y_t$ or $y_E$ is constrained to be larger than about 1.3. Still in \cite{Antola:2010jk} the present authors have shown, by calculating the two loops beta functions of the MSCT couplings, that even for larger values the Yukawa couplings flow to a UV fixed point or decrease asymptotically to zero. The theory therefore allows in principle phenomenologically viable fermion masses while being UV safe.

The scalar sector and other phenomenological constraints are studied in more detail in \cite{Antola:2010jk}.

\subsection{Strong MSCT (sMSCT)}

In this section we study the MSCT model in the regime where it looks like MWT. There are two essential differences from pure MWT. First, the soft supersymmetry breaking\footnote{The most general soft supersymmetry breaking Lagrangian is given in Appendix \ref{appxA}.} parameters introduce additional scales that must have some hierarchy: the technisquarks should decouple from low energy dynamics, while the technigaugino ($\bar{D}_R$) should not. The elementary Higgs scalars and other superpartners in the MSSM sector should not be heavier than the the technisquarks. If the elementary Higgses are very heavy, they can be decoupled and the low energy spectrum is exactly that of MWT. Although we are mainly discussing the situation in which the Higgs masses are positive, one could also study the case where the MSSM Higgses participate actively in EWSB.

Second, the model includes sources of explicit $SU(4)_R$ breaking: the up-type techniquark Yukawa coupling, and the techni-gaugino mass. The latter term explicitly breaks the SU$(2)$ custodial symmetry, which usually guarantees that the electroweak parameter $T$ \cite{Peskin:1990zt} is small in technicolor models. We will instead find a negative contribution that can, however, be offset within MSCT by the new leptons. These terms will also cause the structure of the vevs to be different from pure MWT.

Lifting the soft supersymmetry breaking scale of the MSSM sector, denoted by $\Lambda$, alleviates the flavor-changing and CP breaking problems of the usual MSSM. We now provide a simple estimate for this scale, and therefore for the masses of the MSSM superpartners. In terms of the Yukawa coupling $y_U$ introduced in Eq. (\ref{spmwt}), the vev of the up-type Higgs is given by

\beq
\langle h^0 \rangle =y_U \frac{\langle \bar{U}_R U_L \rangle}{\Lambda^2}\;.
\eeq
Here everything is renormalized at the scale $\Lambda$. The condensates at different scales are related via the anomalous dimension $\gamma$, which is assumed to be constant between the electroweak scale, $v$, and $\Lambda$:

\beq
\langle \bar{U}_R U_L \rangle_\Lambda=(\frac{\Lambda}{4\pi v})^{\gamma}\langle \bar{U}_R U_L \rangle_{4\pi v}=\Lambda^{\gamma}(4\pi v)^{3-{\gamma}}
\eeq
Thus the top mass, renormalized at the scale $\Lambda$, is given by
\beq
m_t\sim(4\pi v) y_Uy_t(\frac{4\pi v}{\Lambda})^{2-\gamma}
\eeq
We assume this form still holds at the electroweak scale and estimate the Yukawa couplings as ${\cal{O}}(1)$.
Thus we obtain an estimate

\beq
\Lambda\sim \left(\frac{4\pi v}{m_t}\right)^{\frac{1}{2-\gamma}}4\pi v\;.
\eeq
Using example values $\gamma=0.5\dots 1.5$ results in $\Lambda\sim 10^4\dots 10^6$ GeV. Assuming $\Lambda$ is very large, the particle spectrum between the electroweak scale and $\Lambda$ becomes that of MWT, but the effective Lagrangian contains four-fermion interactions induced by decoupling the technisquarks and the Higgs. These four fermion interactions can give $\mathcal{O}(100\%)$ corrections to the size of $\gamma$ estimated in pure MWT \cite{Bursa:2010xn,Catterall:2010du,Antipin:2009dz,Pica:2010xq,Pica:2010mt}, exactly as in extended technicolor  \cite{Fukano:2010yv}, justifying the range of $\gamma$ above.

In this large $\Lambda$ scenario, EWSB is solely due to the technicolor sector, and the MSSM Higgs only transmits EWSB to the fermion sector. Since quantum corrections to the Higgs mass are proportional to the largest scale of non-SUSY physics, and the Higgs mass is assumed to be of the same order, those corrections are $\mathcal{O}(1)$ and technicolor therefore solves the little hierarchy problem. Additionally, the $\mu$ problem of the MSSM is relaxed, since $\mu$ is not constrained anymore by the narrow range of values allowing a vev for the Higgses: this clearly opens up the parameter space that is severely constrained in the MSSM \cite{Strumia:2011dv}. Furthermore, heavy masses for the colored scalars and the gluino, as required by the latest experimental lower bounds from the CMS \cite{Khachatryan:2011tk} and ATLAS \cite{Collaboration:2011qk} collaborations, are not problematic in this limit of MSCT.

The MWT has already been shown to pass many experimental tests \cite{Sannino:2004qp,Dietrich:2005jn}, and with the additional Higgs scalars, it should be possible to generate even the mass of the top quark. Thus in this first study of sMSCT we mainly want to find the effects of the explicit $SU(4)_R$ symmetry breaking on the vacuum structure of the theory. This is a salient feature independent of the decoupling scales of the Higgs scalars. To do this task we simply decouple the technisquarks, and build the effective theory at the electroweak scale. The techniquarks form an SU$(4)_R$ fundamental multiplet 
\beq
\eta_{i}=(U_{L},D_{L},\bar{U}_{R},\bar{D}_{R})_{i}\;,
\eeq
transforming under $g\in$SU$(4)_{R}$ as 
$\eta\rightarrow g\eta$.
The low energy effective Lagrangian of MWT is introduced in detail in \cite{Foadi:2007ue}. It is described in terms of the composite field
\beq
M\sim\eta_{a}\eta_{a}^{\text{T}}\;.
\eeq
which transforms under SU$(4)_{R}$ as $M\rightarrow gMg^{\text{T}}$.
The field $M$ is the only technicolor-singlet spinless field made out of
two techniquarks.

From the technicolor perspective, we have a technicolor theory with full ultraviolet completion. Thus the ad hoc $\mathcal{L}_{ETC}$ Lagrangian of \cite{Foadi:2007ue} takes a definite form dictated by the symmetry breaking structure. We apply the spurion technique for this purpose. We begin by finding the spurion fields with hypothetical transformation properties that make the high energy Lagrangian fulfill the largest possible global symmetry. Writing terms in the effective Lagrangian with the lowest number of these spurion fields then encodes the explicit symmetry breaking correctly, if we can assume the explicit symmetry breaking to be small.

The Yukawa coupling of the technifermions to the up-type Higgs can be written in terms of the $\eta$ fields as

\beq
y_{U}\bar{U}_{R}Q_{L}\cdot \tilde{H}_U=\frac{1}{2}\eta^{T}Y\eta,
\eeq
where $Y(\tilde{H}_U)=1/2$, the dot product on the left hand side denotes contraction with $\epsilon_{\alpha\beta}$, and

\begin{eqnarray}
\tilde{H}_U=\left(\begin{array}{c}
h^{+}\\
h^{0}\end{array}\right),\;\;
Y & = &\frac{y_U}{\sqrt{2}}\left(\begin{array}{ccccc}
0 & 0 & h^0 & 0\\
0 & 0 & -h^+ & 0\\
h^0 & -h^+ & 0 & 0\\
0 & 0 & 0 & 0\\
\end{array}\right)
\end{eqnarray}
In order to preserve the SU$(4)$ symmetry the spurion must transform
as $Y\rightarrow g^{*}Yg^{\dagger}$.

We must also take into account the soft SUSY breaking mass of the
$\bar{D}_{R}$ gaugino. We have
\beq
M_{D}\bar{D}_{R}\bar{D}_{R}=\eta^{T}X\eta\;,
\eeq
where
\beq
X=
{\textrm{diag}}(0,0,0,M_D)
\eeq 
transforms under SU$(4)_{R}$ as $X\rightarrow g^{*}Xg^{\dagger}$.

The lowest order contributions to the effective Lagrangian are
\beq
c_{1}v_{w}^{2}\text{Tr}\left[MX\right]+c_{2}v_{w}^{2}\text{Tr}\left[MY\right],
\eeq
where $c_{i}$ are dimensionless unknown low energy constants and $v_{w}$ is
the electroweak scale, arising for dimensional reasons. These terms break the SU$(4)_R$ down to SU$(2)_L\times$ U$(1)_Y$ guaranteeing a mass for all unwanted Goldstone modes.

Thus the full Lagrangian is
\beq
\mathcal{L}_{MSSM}+\frac{1}{2}\text{Tr}\left[DM^{\dagger}DM\right]-\mathcal{V}_{M}
\eeq
where the covariant derivative is introduced in  \cite{Foadi:2007ue} and
\be
\mathcal{V}_{M}= &&-\frac{m_{M}^{2}}{2}\text{Tr}\left[M^{\dagger}M\right]+\frac{\lambda_{M}}{4}\text{Tr}\left[M^{\dagger}M\right]^{2}+\lambda_{M}^{'}\text{Tr}\left[M^{\dagger}MM^{\dagger}M\right]-2\lambda_{M}^{''}\left[\det M+\det M^{\dagger}\right] \nonumber \\
&&-\left(c_1 v_w^2 \text{Tr}\left[MX\right]+c_2 v_w^2 \text{Tr}\left[MY\right]+c.c\right).
\ee

The minimum of the potential determines the vacuum structure. The
full set of minimum equations is given by minimizing  $\cal{V}={\mathcal{V}}_M+{\mathcal{V}}_{MSSM}$ with respect to all scalars,

\beq\label{minVs}
\left\langle \frac{\partial \cal{V}}{\partial\phi_{i}}\right\rangle =0,
\eeq
where $i$ runs over all the scalar fields and each field is set to
its vev after the derivative has been taken. We search for a CP-conserving
electromagnetically neutral vacuum:

\beq
\left\langle M\right\rangle =\frac{1}{2}\left(\begin{array}{cccc}
 0 & 0 & v_{1}+v_{3} & 0\\
 0 & \sqrt{2}\ v_2 & 0  & v_{1}-v_{3}\\
v_{1}+v_{3} & 0 & 0 & 0\\
0 & v_{1}-v_{3} & 0  &\sqrt{2}\ v_4\end{array}\right)\;\;;\;\;\left\langle H_{u}^{0}\right\rangle =\frac{v_u}{\sqrt{2}}\;\;;\;\;\left\langle H_{d}^{0}\right\rangle =\frac{v_d}{\sqrt{2}}
\eeq

The minimum gives us six independent equations, meaning that all the parameters $v_1 ... v_4$, $v_u$, and $v_d$ are needed. By studying the potential, we find that the spurion $X$ tilts the potential in the $v_{4}$ direction, forcing $v_4$ and also $v_2$ to be nonzero. In other words, either all $v_2$, $v_4$ and $M_D$ are zero, or they are all nonzero.

To give an explicit but simple solution to the minimum equations (\ref{minVs}), we note that if $M_D$ is small, we can consider the limit $v_2=v_4=0$ as an approximate solution. Therefore we set $M_D=v_2=v_4=0$, and solve Eq.(\ref{minVs}) with respect to the remaining vevs. In the limit of large Higgs squared masses and small $v_H=\sqrt{v_u^2+v_d^2}$ we find
\beq
v_3=v_1,\ v_H=\frac{2 c_1 y_U v_w^2 v_1}{b^2-(m^2_d+\mu^2) (m^2_u+\mu^2)}\sqrt{b^2+\left(m^2_d+\mu^2\right)^2}.
\eeq 

It is clear the non-zero vev of the MSSM Higgses is solely caused by the linear interaction term with the technimeson field $M$. Moreover, the value of $v_H$ is roughly proportional to the ratio $v_w/m_H$ (when $v_1\sim v_w$), and therefore for an extremely heavy Higgs scalar we find the expected behavior, with the vev $v_H$ becoming negligible.

The oblique corrections $S$ and $T$ \cite{Peskin:1990zt} are defined as
\be
\alpha S &=&4 s_w^2c_w^2\frac{\Pi^{new}_{ZZ}(m_Z^2)-\Pi^{new}_{ZZ}(0)}{m_Z^2},\nonumber \\
\alpha T &=& \frac{\Pi^{new}_{WW}(0)}{m^2_W}-\frac{\Pi^{new}_{ZZ}(0)}{m^2_Z},
\ee
where $\Pi^{new}_{ij}$ are self-energies. The label {\it new} is a reminder that the origin of the $(S,T)$-plane actually corresponds to the SM with a reference value of the Higgs mass, denoted by $m_{\textrm{ref}}$. Since we have removed the SM Higgs sector and added the technicolor and MSSM Higgs sectors, the correct $T$-parameter is given by
\beq
T=T_{\textrm{SM}}(m_{\textrm{ref}})-T_H(m_{\textrm{ref}})+T_{\textrm{full}}=T_{\textrm{full}}-T_H(m_{\textrm{ref}}),
\eeq
because $T_{\textrm{SM}}(m_{\textrm{ref}})=0$ by definition. Here $T_H$ denotes contributions from the SM Higgs sector and $T_{full}$ is calculated by including all the contributions in the model. Similar formulas hold for $S$. For $T$, we need the eletroweak gauge boson masses:
\be
m_{W^{\pm}}^{2} &=& \frac{g_{W}^{2}}{4}(v_{1}^{2}+v_{2}^{2}+v_3^2+v_{u}^{2}+v_{d}^{2}),\nonumber \\
m_{Z}^{2} &=& \frac{g_{W}^{2}+g_{Y}^{2}}{4}(v_{1}^{2}+2v_{2}^{2}+v_3^2+v_{u}^{2}+v_{d}^{2}).
\ee

These lead to a tree-level contribution to $T$:
\begin{eqnarray}
\alpha T_{\textrm{tree}} =-\frac{v_{2}^{2}}{v_{w}^{2}}. \label{T_contrib}
\end{eqnarray}

By using the most recent experimental data \cite{Nakamura:2010zzi}, $T=0.03\pm 0.11$, we find
\beq
|v_2|< 7\ \rm{GeV}.
\eeq

This limit does not take into account the one-loop perturbative contribution of the new leptons. These give generically a positive contribution to $T$ \cite{He:2001tp},

\beq
T_\textrm{lepton}=\frac{1}{8\pi s_w^2c_w^2m_Z^2}\left(\frac{m_E^2+m_N^2}{2}-\frac{m_E^2 m_N^2}{m_E^2-m_M^2}\ln \frac{m_E^2}{m_N^2}\right).
\eeq
This can offset the contribution (\ref{T_contrib}). For example considering $m_E=550$ GeV,  $m_N= 470$ GeV, and $v_2=7$ GeV gives $T_{\textrm{tree}}+T_{\textrm{lepton}}=0.005$.

\begin{figure}[h]
\begin{center}
\includegraphics[width=18pc]{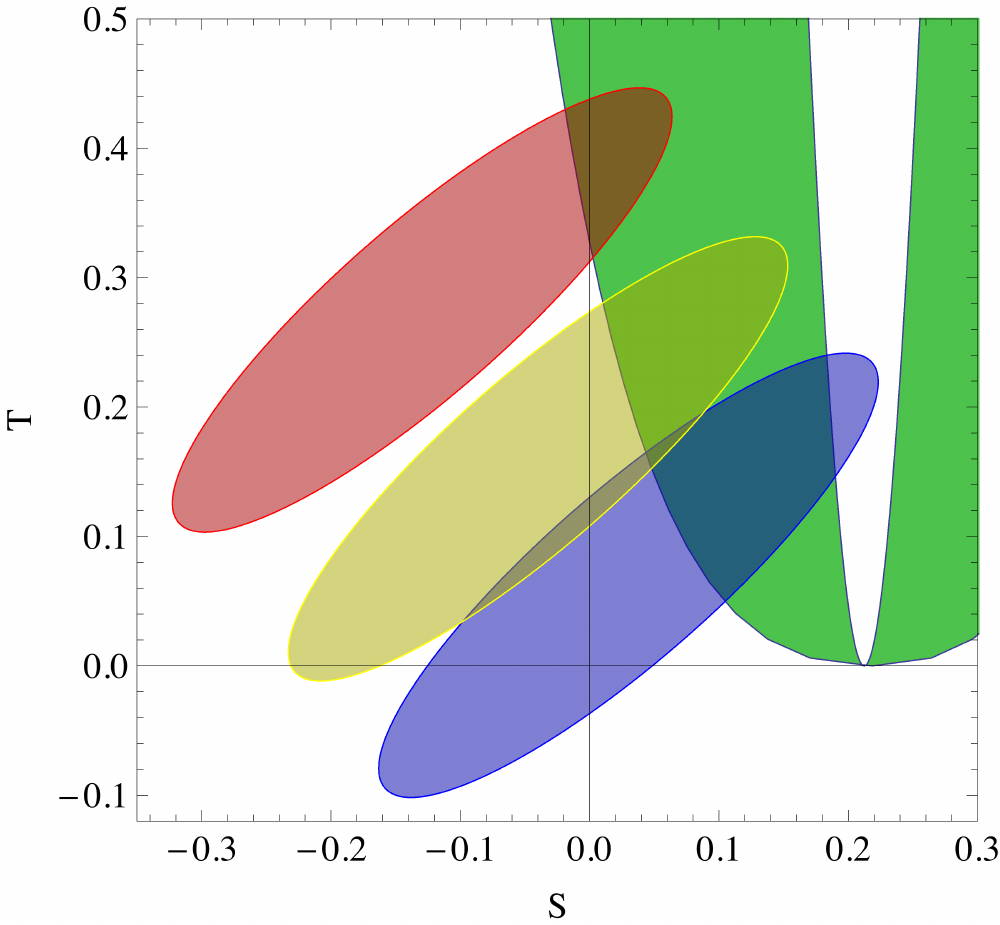}
\caption{\label{viaST}The ellipses represent the 90\% confidence region for the S and T parameters, and are obtained, from lower to higher, for a reference Higgs mass of 117 GeV, 300 GeV, and 1 TeV, respectively. The contribution from the MSCT theory as function of the $N$ and $E$ lepton masses is expressed by the green region, with $m_Z \leqslant m_{N,E}\leqslant 10 m_Z$ and assuming $v_2=0$.}
\vspace{1pc}
\end{center}
\end{figure}

The intrinsic contribution of the technicolor sector to the S parameter has been estimated by calculating the one loop contribution of heavy techniquarks. This so-called naive estimate gives for MWT $S_{\mathrm{naive}}\simeq 1/2\pi$. The new leptons also give a perturbative contribution to $S$ \cite{He:2001tp}. It is interesting to note, that with all these contributions combined one can actually fit the electroweak data very well for any value of the lightest Higgs mass. This is seen from Fig. \ref{viaST}, where we have not included the effects of $v_2$ which would offset the base point of the parabola to negative values of T.

\section{Minimal Supersymmetric Technicolor (MST)}
\label{n1model}
A more straightforward supersymmetrization of the MWT can be obtained simply adding a superpartner for each particle in Eq.(\ref{MWTp}) and for the techni-gluon $G$. We call the resulting model Minimal Supersymmetric Technicolor (MST). This model features the most economical supersymmetric extension of MWT for the generic anomaly-free hypercharge assignment of Eq.(\ref{assign2}). The only new Weyl fermion among the techni-superpartners is the techni-gaugino, and it does not introduce new anomalies.


We define the chiral superfields: 
\beq
\left(\tilde{U}_L,\ U_L\right)\in \Phi_1,\quad \left(\tilde{D}_L,\ D_L\right)\in \Phi_2,\quad   
\left(\tilde{\bar{U}}_R,\ \bar{U}_R\right)\in U,\quad\left(\tilde{\bar{D}}_R,\ \bar{D}_R\right)\in D,  
\label{superq2}
\eeq
all transforming according to the adjoint representation to the $SU(2)_{TC}$ gauge group, and the gauge superfield
\beq
\left(G,\ \lambda\right)\in V \ .
\eeq
The heavy lepton superfields are defined in Eq.(\ref{superl}).


We choose to introduce in the theory the two Higgs superfields whose charges are defined in Eq.(\ref{superH}), though for some value of $y$ this is not the minimal choice\footnote{For $y=-\frac{1}{3}$ the heavy lepton $\tilde{E}_L$ and the neutrino $\tilde{\nu}_L$ could play the role of $\tilde{H}_2$ and $\tilde{H}_1^\prime$, respectively, even though constraints on L number violating processes require the corresponding Yukawa couplings to be very small.}. The superpotential of the theory is also dictated by the value of the hypercharge parameter $y$ in Eq.(\ref{assign2}), since the gauge invariance of a generic term in the superpotential depends on the hypercharge assignment. We find the models obtained for $y=\pm 1,\pm\frac{1}{3}$, particularly appealing, since for these values of $y$ it is possible to write mass and Yukawa terms involving only the superfields in Eq.(\ref{superl}) and Eq.(\ref{superq2}), which are less constrained by the experiment than new terms involving also SM fields.

\subsection{The MST Superpotential for $y=1$}
\label{mst}

Requiring gauge invariance as well as B and L numbers conservation for $y=1$ we find the superpotential
\be
P_{TC}=&-&\frac{g_{TC}}{\sqrt{2}} \epsilon_{ij} \epsilon^{abc} \Phi^a_i \Phi^b_j U^c+y_U \epsilon_{ij}\Phi^a_i H_j U^a+y_D \epsilon_{ij}\Phi^a_i H^\prime_j D^a+y_N \epsilon_{ij3}\Lambda_i H_j N\nonumber\\
&+&y_E \epsilon_{ij}\Lambda_i H^{\prime}_j E+y_R U^a U^a E\,,
\label{spmst1}
\ee
where $a=1,2,3$ is the $SU(2)_{TC}$ gauge index and where we suppressed the family index $k$, as we do for the rest of the paper.

In contrast to the pMSCT case, $D$ now naturally acquires a mass term. We expect therefore the parameter space of the $y=1$ MST to be less constrained by experiment than that of the MSCT. The superpotential for $y=-1$ is similar to the $y=1$ case and we do not present it here.

\subsection{The MST Superpotential for $y=\pm 1/3$}

For $y=\pm\frac{1}{3}$ the MST features both a gauge singlet and a Higgs candidate (with corresponding hypercharges $\pm\frac{1}{2}$): the gauge singlet can be used to solve the MSSM $\mu$ problem in an NMSSM fashion, while the Higgs candidate can be used in principle to reduce further the particle content of the theory. Indeed for $y=-\frac{1}{3}$ we could identify the Higgs superfields $H$ with $\Lambda$ and $H^{\prime}$ with $l$ (a generic MSSM leptonic weak doublet superfield). The experiment, though, constraints the value of the Yukawa couplings to such Higgs scalars to be very small, since the corresponding terms violate $L$ number. The resulting mass spectrum would be too light to be viable. We will therefore present the model with $y=-\frac{1}{3}$ which includes both the $H$ and $H^{\prime}$.

We start with the hypercharge assignment given by $y=\frac{1}{3}$, that corresponds to that of a SM family (assuming that that includes also a right-handed neutrino):
\be
Y(Q_L)&=&\frac{1}{6} \ ,\qquad Y(\bar{U}_R,\bar{D}_R) =\left(-\frac{2}{3},\frac{1}{3}\right) \ ,  \nonumber\\
Y(L_L)&=&-\frac{1}{2} \ ,\qquad Y(\bar{N}_R,\bar{E}_R) = \left(0,1\right) \ \label{assign3} .
\ee
Following the notation of Eq.(\ref{spot}) we write the extension of the MSSM superpotential as
\be
P_{TC}&=&s_N N+\frac{1}{2}m_N N N+m_{\Lambda}\epsilon_{ij}\Lambda_i H_j+y_N N^3+y_U \epsilon_{ij} \Phi_i^a H_j U^a+y_D \epsilon_{ij} \Phi_i^a H^{\prime}_j D^a\nonumber\\
&+&y'_N \epsilon_{ij} \Lambda_i H_j N+y_E \epsilon_{ij} \Lambda_i H^{\prime}_j E+y'_D \epsilon_{ij} \Phi^a_i \Lambda_j D^a+y_H \epsilon_{ij} H_i H^{\prime}_j N+y_d \epsilon_{ij} q_i \Lambda_j d,\nonumber\\
&+&y_e \epsilon_{ij} l_i \Lambda_j e
\label{spoN1}
\ee
where $q_i$ is the chiral superfield associated with the $i$-th component of the SM quark weak doublet (the family index $k$ has been suppressed), while the chiral superfield $d$ contains the SM $\bar{d}_R$ quark. We can add to the previous equation the lepton number violating terms\footnote{We consider all the techni-superfields to have baryon and lepton number equal to zero.}
\beq
P_{TC,\Delta L\neq 0}=y'_E \epsilon_{ij} \Lambda_i l_j E+y''_D \epsilon_{ij} \Phi^a_i l_j D^a+y_n \epsilon_{ij} l_i H_j N+y'_e \epsilon_{ij} l_i H^{\prime}_j E+y''_e \epsilon_{ij} \Lambda_i H^{\prime}_j e.
\label{spoN1lv}
\eeq
Notice that the term proportional to $m_N$ in Eq.(\ref{spoN1}) and that proportional to $y_n$ in Eq.(\ref{spoN1lv}) generate the terms required to give mass to the neutrino in a natural way by the seesaw mechanism, allowing $y_n$ to be of the same order as the other Yukawa coupling constants.

The hypercharge assignment for $y=-\frac{1}{3}$ is equal to minus that of a SM family (including a right-handed neutrino):
\be
Y(Q_L)&=&-\frac{1}{6} \ ,\qquad Y(\bar{U}_R,\bar{D}_R) =\left(-\frac{1}{3},\frac{2}{3}\right) \ ,  \nonumber\\
Y(L_L)&=&\frac{1}{2} \ ,\qquad Y(\bar{N}_R,\bar{E}_R) = \left(-1,0\right) \ \label{assign4} .
\ee
The corresponding $B$ and $L$ number conserving superpotential is
\be
P_{TC}&=&s_E E+\frac{1}{2}m_E E E+m_{\Lambda} \epsilon_{ij}{\Lambda}_i H^{\prime}_j+y_E E^3+y_{D} \epsilon_{ij} {\Phi}_i^a H^{\prime}_j {D}^a+y_{U} \epsilon_{ij} {\Phi}_i^a H_j {U}^a\nonumber\\
&+&y'_E \epsilon_{ij} {\Lambda}_i H^{\prime}_j E+y_{N} \epsilon_{ij} {\Lambda}_i H_j {N}+y'_{U} \epsilon_{ij} {\Phi}^a_i {\Lambda}_j {U}^a+y_H \epsilon_{ij} H_i H^{\prime}_j E+y_u \epsilon_{ij} q_i \Lambda_j u,
\label{spoN2}
\ee
where the fermionic component of the chiral superfield $u$ is the SM quark $\bar{u}_R$.

The lepton number violating terms include also mass-mixing terms obtained coupling techni-singlet and MSSM leptonic superfields with opposite hypercharges. These are in addition to a number of Yukawa terms. The latters arise as a direct consequence of $l$ and ${\Lambda}$ having the same charge assignments as $H^{\prime}$ and $H$. We can now add to $P_{MSSM}$ also the superpotential
\be
P_{TC,\Delta L\neq 0}&=&m'_{\Lambda}\epsilon_{ij} \Lambda_i l_j+m_e {N} e+y'_{D}\epsilon_{ij}{\Phi}^a_i l_j D^a+y''_{E}\epsilon_{ij}{\Lambda}_i l_j E+y_n\epsilon_{ij} l_i H_j E \nonumber\\
&+&y'_n\epsilon_{ij} l_i {\Lambda}_j E+ y_e N e E, 
\ee
where the terms proportional to $y_n$ and $y'_n$ allow, together with that proportional to $m_E$ in Eq.(\ref{spoN2}), to solve the neutrino mass naturalness problem by the seesaw mechanism.

\section{Conclusions}
We have presented novel extensions of the SM featuring an ${\cal N}=4 $ or ${\cal N}=1$ supersymmetric technicolor sector. These models are minimal and direct supersymmetric generalizations of the MWT model.
 
We started from the observation that the MWT model has the same degrees of freedom as the ${\cal N}=4$ supermultiplet, except for the absence of the six real scalars. Following this trail we added the six scalars and constructed an extension of the SM naturally featuring a supersymmetrized version of MWT. In the MSSM we then embedded the ${\cal N}=4$ supersymmetric technicolor sector in such a way that the extended SUSY is broken to ${\cal N}=1$ only via EW gauge and Yukawa interactions. Since the original MWT model contains also a natural 4th family of leptons, needed to cure the topological Witten anomaly, we introduced in the theory also a 4th family of lepton superfields. We then constructed the superpotential for the full theory and provided the Lagrangian in terms of superfields as well as the corresponding elementary field components. The resulting model was termed in short MSCT.

Depending on the way SUSY breaks, the value of the technicolor coupling constant around the EW scale, and the value assumed by several other couplings, one may investigate several different physical scenarios ranging from ordinary technicolor to unparticle models as well as perturbative extensions. We considered the basic features of the perturbative and the technicolor-like regimes of MSCT. We found that in the perturbative regime, the Yukawa couplings are required to be large. This though does not represent a problem since the present authors have shown in \cite{Antola:2010jk}, by calculating the beta functions at two loops, that the the Yukawa couplings flow either to a UV fixed point or to zero, and therefore that MSCT is UV safe. In the strong regime, we found a tree-level contribution to $T$, that can be offset by the perturbative one-loop contribution of the new leptons.

Since the new sector coupled to the MSSM is conformal, one can use AdS/CFT methods when the supersymmetric technicolor coupling constant is taken to be large. Besides, the model can benefit from, and provide motivation for, lattice studies of ${\cal N}=4$ SUSY (see \cite{Catterall:2005fd,Elliott:2008jp,Giedt:2008xm} for recent interesting lattice investigations). 

For completeness we have also considered the case in which the MWT supersymmetric technicolor extension is directly an ${\cal N}=1$ gauge theory, the MST. Here more fields than in the case of the MSCT are needed.  We constructed the superpotential for several choices of the hypercharges of the technifields. 

\acknowledgments
We would like to thank Matti J\"arvinen for useful discussions. 

\newpage
\appendix

\section{${\cal N}=4$ Super Yang-Mills: Notation and Lagrangian}
\label{N4susyl}
The ${\cal N}=4$ supersymmetric Lagrangian for an $SU(N)$ gauge theory can be written in terms of three ${\cal N}=1$ chiral superfields $\Phi_i,\ i=1,2,3$ and one ${\cal N}=1$ vector superfield $V$, all in the adjoint representation of $SU(N)$. The superpotential for this Lagrangian reads (see \cite{Dorey:2000fc} and references therein)
\beq
P=-\frac{g}{3\sqrt{2}} \epsilon_{ijk} f^{abc} \Phi^a_i \Phi^b_j \Phi^c_k,\ i,j,k=1,2,3; a,b,c=1,\cdots,N^2-1;
\label{sp4}
\eeq
where $g$ is the gauge coupling constant, and $f^{abc}$ the structure constant. This superpotential is invariant under $SU(3)$ transformations over the flavor index. The full Lagrangian is indeed invariant under $SU(4)$ transformations because the ${\cal N}=4$ SUSY algebra is invariant under the same transformations of the supercharges.

Following the notation of Wess and Bagger \cite{Wess:1992cp} we write
\beq
{\cal L} = \frac{1}{2} \tr \left(W^{\alpha} W_{\alpha}|_{\theta\theta}+\bar{W}_{\dot{\alpha}} \bar{W}^{\dot{\alpha}}|_{\bar{\theta}\bar{\theta}}\right)+\Phi_i^{\dagger}\exp \left( 2 g V \right) \Phi_i|_{\theta\theta\bar{\theta}\bar{\theta}}+\left(P|_{\theta\theta}+h.c.\right)
\label{LPhi}
\eeq
where 
\beq
W_\alpha=-\frac{1}{4 g}\bar{D}\bar{D}\exp\left(-2 g V\right)D_\alpha\exp\left(2 g V\right),\ \ V=V^a T^a_A,\ \  \left(T^a_A\right)^{bc}=-i f^{abc},
\eeq
and with $\Phi_i$ having gauge components $\Phi_i^a$. In terms of the component fields Eq.(\ref{LPhi}) can be expressed as
\be
{\cal L}&=&-\frac{1}{4} F^{\mu\nu a}F_{\mu\nu}^a-i \bar{\lambda}^a\bar{\sigma}^\mu D_\mu \lambda^a-D^\mu \phi^{a\dagger}_i D_\mu \phi^a_i - i \bar{\psi}^a_i\bar{\sigma}^\mu D_\mu\psi^a_i \nonumber\\
&&+\sqrt{2} g f^{abc} \left( \phi_i^{a\dagger} \psi_i^b\lambda^c+\bar{\lambda}^c\bar{\psi}_i^b \phi_i^a \right) +\frac{g}{\sqrt{2}}\epsilon_{ijk} f^{abc}\left( \phi_i^a\psi_j^b\psi^c_k+\bar{\psi}_k^c\bar{\psi}_j^b \phi_i^{a\dagger} \right)\nonumber\\
&&- \frac{1}{2} g^2 \left( f^{abd} f^{ace} + f^{abe}f^{acd}  \right)\phi_i^{b\dagger} \phi_i^c  \phi_j^{d\dagger} \phi_j^e
\label{Lphi}
\ee
where
\beq
F^{a}_{\mu\nu}=\partial_\mu A^a_\nu-\partial_\nu A^a_\mu- g f^{abc} A^b_\mu A^c_\nu,\ D_\mu \xi^a=\partial\xi^a-g f^{abc} A^b_\mu \xi^c,\ \xi=\lambda,\psi_i,\phi_i.
\eeq
Here $\lambda$ is the gaugino, while $\psi_i$ and $\phi_i$ are respectively the fermionic and scalar component of $\Phi_i$.
To make explicit the $SU(4)$ R-symmetry of the Lagrangian the following change of
variables provides useful:
\beq
\varphi^a_{rs}=-\varphi^a_{sr},\ \varphi^a_{i4}=\frac{1}{2}\phi^a_i,\ \varphi^{a}_{ij}=\frac{1}{2}\epsilon_{ijk}\phi^{a\dagger}_k,\ \eta^a_i=\psi^a_i,\ \eta^a_4=\lambda^a;\ r,s=1,\cdots,4.
\eeq
The symmetry of the Lagrangian can be made manifest by rewriting Eq.(\ref{Lphi}) as
\be
{\cal L}&=&-\frac{1}{4} F^{\mu\nu a}F_{\mu\nu}^a-\tr D^\mu \varphi^{a\dagger} D_\mu \varphi^a - i \bar{\eta}^a_r\bar{\sigma}^\mu D_\mu\eta^a_r \nonumber\\
&& -\sqrt{2}g f^{abc}\left( \varphi_{rs}^{a\dagger}\eta_r^b\eta^c_s+\bar{\eta}_r^c \bar{\eta}^b_s \varphi_{rs}^{a} \right)\nonumber\\
&&- \frac{1}{2} g^2 \left(f^{abd} f^{ace} + f^{abe}f^{acd} \right)\tr\varphi^{b\dagger} \varphi^c \tr  \varphi^{d\dagger} \varphi^e.
\label{Lvarphi}
\ee
Under $SU(4)$ $\varphi^a$ transforms as a \textbf{6}, $\eta^a$ as a \textbf{4}, and $A^a_\mu$ as a \textbf{1}, leaving the Lagrangian in Eq.(\ref{Lvarphi}) unchanged.

\section{MSCT Lagrangian}
\label{appxA}
The Lagrangian of a supersymmetric theory can, in general, be defined by
\beq
{\cal L}={\cal L}_{kin}+{\cal L}_{g-Yuk}+{\cal L}_{D}+{\cal L}_{F}+{\cal L}_{P-Yuk}+{\cal L}_{soft},
\label{LP}
\eeq
where the labels refer to the kinetic terms, the Yukawa ones given by gauge and superpotential interactions, the $D$ and $F$ scalar interaction terms, and the soft SUSY breaking ones. All these terms can be expressed in function of the elementary fields of the theory with the help of the following equations:
\be
{\cal L}_{kin}&=&-\frac{1}{4} F_j^{\mu\nu a}F_{j\mu\nu}^a-i \bar{\lambda}_j^a\bar{\sigma}^\mu D_\mu \lambda_j^a-D^\mu \phi^{a\dagger}_i D_\mu \phi^a_i - i \bar{\chi}^a_i\bar{\sigma}^\mu D_\mu\chi^a_i\ ,\label{comp1}\\
{\cal L}_{g-Yuk}&=&\sum_j i\sqrt{2} g_j \left( \phi_i^{\dagger} T^a_j\chi_i\lambda^a_j-\bar{\lambda}^a_j\bar{\chi}_i T^a_j \phi_i \right)\ ,\label{comp2}\\
{\cal L}_{D}&=&-\frac{1}{2}  \sum_j g_j^2 \left(\phi_i^{\dagger} T^a_j \phi_i \right)^2\ ,\label{comp3}\\
{\cal L}_{F}&=&-\left|\frac{\partial P}{\partial \phi^a_i} \right|^2\ ,\label{comp4}\\
{\cal L}_{P-Yuk}&=&-\frac{1}{2}\left[\frac{\partial^2 P}{\partial \phi^a_i\partial \phi^b_l}\chi^a_i\chi^b_l+h.c. \right]\label{comp5},
\ee
where $i,l$ run over all the scalar field labels, while $j$ runs over all the gauge group labels, and $a,b$ are the corresponding gauge group indices. Furthermore, we normalize the generators in the usual way, by taking the index $T(F)=\frac{1}{2}$, where 
\[
\tr T^a_RT^b_R=T(R)\delta^{ab},
\]
with $R$ here referring to the representation ($F$=fundamental).
The SUSY breaking soft terms, moreover, are obtained by re-writing the superpotential in function of the scalar fields alone, and by adding to it its Hermitian conjugate and the mass terms for the gauginos and the scalar fields.

We refer to \cite{Martin:1997ns} and references therein for the explicit form of ${\cal L}_{MSSM}$ in terms of the elementary fields of the MSSM, and focus here only on ${\cal L}_{TC}$. The kinetic terms are trivial and therefore we do not write them here. The gauge Yukawa terms are given by
\be
{\cal L}_{g-Yuk}&=& \sqrt{2}g_{TC}\left(  \tilde{\bar{U}}_L^bU_L^c\bar{D}^a_R-D_R^a\bar{U}_L^b\tilde{U}_L^c + \tilde{\bar{D}}_L^bD_L^c\bar{D}^a_R-D_R^a\bar{D}_L^b\tilde{D}_L^c
+ \tilde{U}_R^b\bar{U}_R^c\bar{D}^a_R-D_R^aU_R^b\tilde{\bar{U}}_R^c  \right)\epsilon^{abc}\nonumber\\
&+&i\frac{g_L}{\sqrt{2}} \left(  \tilde{\bar{Q}}_L^i Q_L^j\tilde{W}^k-\tilde{\bar{W}}^k \bar{Q}_L^i \tilde{Q}_L^j+\tilde{\bar{L}}_L^i L_L^j\tilde{W}^k-\tilde{\bar{W}}^k \bar{L}_L^i \tilde{L}_L^j  \right)\sigma^k_{ij}\nonumber\\
&+&i\sqrt{2} g_Y\sum_p Y_p\left(  \tilde{\bar{\chi}}_p\chi_p\tilde{B}-\tilde{\bar{B}}\bar{\chi}_p\tilde{\chi}_p  \right),\ \chi_p=U^a_L,D^a_L,\bar{U}^a_R,N_L,E_L,\bar{N}_R,\bar{E}_R\ ,
\ee
where $\tilde{W}^k$ and $\tilde{B}$ are respectively the wino and the bino, $\sigma^k$ the Pauli matrices, $i,j=1,2;\ k,a,b,c=1,2,3$; and the hypercharge $Y_p$ is given for each field $\chi_p$ in Table \ref{MSCTsuperfields}.

The $D$ terms are given by
\beq
{\cal L}_D=-\frac{1}{2}\left( g_{TC}^2 D^a_{TC} D^a_{TC}+ g_{L}^2 D^k_{L} D^k_{L}+ g_{Y}^2 D_{Y} D_{Y}\right)+\frac{1}{2}\left( g_{L}^2 D^k_{L} D^k_{L}+ g_{Y}^2 D_{Y} D_{Y}\right)_{MSSM},
\eeq
where
\be
D^a_{TC}&=&-i \epsilon^{abc}\left(  \tilde{\bar{U}}_L^b\tilde{U}^c_L+ \tilde{\bar{D}}_L^b\tilde{D}^c_L+ \tilde{U}_R^b\tilde{\bar{U}}^c_R  \right),\ D^k_{L}=\frac{\sigma^k_{ij}}{2} \left(  \tilde{\bar{Q}}_L^{i\,a}\tilde{Q}^{j\,a}_L + \tilde{\bar{L}}_L^i\tilde{L}^j_L  \right) +D^k_{L,MSSM}\nonumber\\
D_{Y}&=&\sum_p Y_p \tilde{\bar{\chi}}_p\tilde{\chi}_p+D_{Y,MSSM}.
\ee
In these equations the $D^k_{L, MSSM}$ and $D_{Y, MSSM}$ auxiliary fields are assumed to be expressed in function of the MSSM elementary fields \cite{Martin:1997ns}.
The rest of the scalar interaction terms\footnote{We consider the constants in the superpotential to be real to avoid the contribution of CP violating terms.} is given by
\be
{\cal L}_F&=&-g_{TC}^2\left[ \left(  \tilde{U}_L^b\tilde{\bar{U}}_L^b+\tilde{D}_L^b\tilde{\bar{D}}_L^b+\tilde{\bar{U}}_R^b\tilde{U}_R^b\right)^2- \left(  \tilde{U}_L^b\tilde{\bar{U}}_L^c+\tilde{D}_L^b\tilde{\bar{D}}_L^c+\tilde{\bar{U}}_R^b\tilde{U}_R^c\right)\left(  \tilde{\bar{U}}_L^b\tilde{U}_L^c+\tilde{\bar{D}}_L^b\tilde{D}_L^c\right.\right.\nonumber\\
&+&\left.\left.\tilde{U}_R^b\tilde{\bar{U}}_R^c\right) \right]-y_U^2\left[  \left(\tilde{H}_1\tilde{D}_L^a-\tilde{H}_2\tilde{U}_L^a\right)\left(\tilde{\bar{H}}_1\tilde{\bar{D}}_L^a-\tilde{\bar{H}}_2\tilde{\bar{U}}_L^a\right) +\tilde{U}_R^a\tilde{\bar{U}}_R^a\left(\tilde{H}_1\tilde{\bar{H}}_1+\tilde{H}_2\tilde{\bar{H}}_2\right)\right.\nonumber\\
&+&\left.\tilde{U}_R^a\tilde{\bar{U}}_R^b\left(\tilde{\bar{U}}_L^a\tilde{U}^b_L+\tilde{\bar{D}}_L^a\tilde{D}_L^b\right)  \right]-y^2_N\left[ \left( \tilde{\bar{N}}_L \tilde{\bar{H}}_2 -\tilde{\bar{E}}_L \tilde{\bar{H}}_1 \right)\left( \tilde{N}_L \tilde{H}_2 -\tilde{E}_L \tilde{H}_1 \right)\right.\nonumber\\
&+&\left.\tilde{N}_R\tilde{\bar{N}}_R\left(\tilde{H}_1\tilde{\bar{H}}_1+\tilde{H}_2\tilde{\bar{H}}_2+\tilde{N}_L\tilde{\bar{N}}_L+\tilde{E}_L\tilde{\bar{E}}_L\right) \right]
-y^2_E\left[ \left( \tilde{\bar{N}}_L \tilde{\bar{H}}^{\prime}_2 -\tilde{\bar{E}}_L \tilde{\bar{H}}^{\prime}_1 \right)\left( \tilde{N}_L \tilde{H}^{\prime}_2 -\tilde{E}_L \tilde{H}^{\prime}_1 \right)\right.\nonumber\\
&+&\left.\tilde{E}_R\tilde{\bar{E}}_R\left(\tilde{H}^{\prime}_1\tilde{\bar{H}}^{\prime}_1+\tilde{H}^{\prime}_2\tilde{\bar{H}}^{\prime}_2+\tilde{N}_L\tilde{\bar{N}}_L+\tilde{E}_L\tilde{\bar{E}}_L\right) \right]
-y_R^2\left( \tilde{U}_R^a \tilde{U}_R^a \tilde{\bar{U}}_R^b \tilde{\bar{U}}_R^b+4 \tilde{\bar{U}}_R^a \tilde{U}_R^a \tilde{\bar{E}}_R \tilde{E}_R\right) \nonumber\\
&+&\left\{ \sqrt{2}y_U g_{TC} \epsilon^{abc}\left[\tilde{U}_L^b\tilde{D}_L^c\left(\tilde{\bar{H}}_1\tilde{\bar{D}}^a_L-\tilde{\bar{H}}_2\tilde{\bar{U}}^a_L\right) \right. +\tilde{U}^a_R\tilde{\bar{U}}_R^b\left(\tilde{U}^c_L\tilde{\bar{H}}_1+\tilde{D}^c_L\tilde{\bar{H}}_2\right)\right] \nonumber\\
&-&y_U y_N \tilde{U}^a_R\tilde{\bar{N}}_R \left(\tilde{\bar{U}}^a_L\tilde{N}_L+\tilde{\bar{D}}^a_L\tilde{E}_L\right)-
y_N y_E\left.\tilde{N}_R\tilde{\bar{E}}_R\left( \tilde{\bar{H}}_1\tilde{H}^{\prime}_1+\tilde{\bar{H}}_2\tilde{H}^{\prime}_2 \right)\right. \nonumber\\
&+&y_R \tilde{\bar{U}}_R^a \left[2 \sqrt{2}g_{TC} \epsilon^{abc}\tilde{\bar{U}}_L^b \tilde{\bar{D}}_L^c \tilde{\bar{E}}_R+2 y_U \tilde{\bar{E}}_R \left( \tilde{\bar{D}}^a_L \tilde{\bar{H}}_1 -  \tilde{\bar{U}}^a_L\tilde{\bar{H}}_2 \right)
+y_E \tilde{\bar{U}}_R^a \left( \tilde{\bar{E}}_L \tilde{\bar{H}}_1^\prime- \tilde{\bar{N}}_L \tilde{\bar{H}}_2^\prime \right) \right]
\nonumber\\
&+&\left.h.c.\right\}+{\cal L}_{mix},
\label{LF4}
\ee
with ${\cal L}_{mix}$ defined in function of the $F$ auxiliary fields associated with the MSSM two Higgs super-doublets:
\be
{\cal L}_{mix}&=&-\sum_{\phi_p} \left(F_{\phi_p,TC} F^\dagger_{\phi_p,MSSM}+h.c.\right),\ \phi_p=H^\prime_1,H^\prime_2,H_1,H_1,\ 
F_{H^\prime_1,TC}=-y_E\tilde{E}_L\tilde{\bar{E}}_R,\nonumber\\ 
F_{H^\prime_2,TC}&=&y_E\tilde{N}_L\tilde{\bar{E}}_R,\ F_{H_1,TC}=-y_U\tilde{D}_L^a\tilde{\bar{U}}_R^a-y_N\tilde{E}_L\tilde{\bar{N}}_R,\ F_{H_2,TC}=y_U\tilde{U}_L^a\tilde{\bar{U}}_R^a+y_N\tilde{N}_L\tilde{\bar{N}}_R.\nonumber\\ 
\label{Lmix}
\ee
The corresponding MSSM auxiliary fields $F$ can be found in \cite{Martin:1997ns} and references therein. Also, in the Eqs.(\ref{LF4},\ref{Lmix}) we used $\tilde{H}$ and $\tilde{H}^{\prime}$ to indicate the scalar Higgs  doublets, for consistency with the rest of the notation where the tilde identifies the scalar component of a chiral superfield or the fermionic component of a vector superfield. The remaining Yukawa interaction terms are determined by the superpotential, and can be expressed as
\be
{\cal L}_{P-Yuk}&=&\sqrt{2}g_{TC}\epsilon^{abc}\left(U_L^a D_L^b\tilde{\bar{U}}_R^c+U_L^a \tilde{D}_L^b\bar{U}_R^c+\tilde{U}_L^a D_L^b\bar{U}_R^c\right) +  y_U  \left[\left( H_1 D_L^a -H_2U_L^a\right)\tilde{\bar{U}}_R^a\right.\nonumber\\
&+&\left.\left( \tilde{H}_1 D_L^a -\tilde{H}_2U_L^a\right)\bar{U}_R^a+\left( H_1 \tilde{D}_L^a -H_2\tilde{U}_L^a\right)\bar{U}_R^a\right]+y_N  \left[\left( H_1 E_L -H_2N_L\right)\tilde{\bar{N}}_R\right.     \nonumber\\
&+&\left.\left( H_1 \tilde{E}_L -H_2\tilde{N}_L\right)\bar{N}_R+\left( \tilde{H}_1 E_L -\tilde{H}_2N_L\right)\bar{N}_R\right]+y_E  \left[\left( H^{\prime}_1 E_L -H^{\prime}_2N_L\right)\tilde{\bar{E}}_R\right.  \nonumber\\
&+&\left.\left( H^{\prime}_1 \tilde{E}_L -H^{\prime}_2\tilde{N}_L\right)\bar{E}_R+\left( \tilde{H}^{\prime}_1 E_L -\tilde{H}^{\prime}_2N_L\right)\bar{E}_R\right]-y_R \bar{U}^a_R \left( \bar{U}_R^a \tilde{\bar{E}}_R+ \bar{\tilde{U}}_R^a \bar{E}_R \right)  
\nonumber\\
&+&h.c..
\ee
The soft SUSY breaking terms, finally, can be written straightforwardly starting from the superpotential in Eq.(\ref{spmwt}), to which we add the techni-gaugino and scalar mass terms as well:
\be
\label{Lsof}
{\cal L}_{soft}&=&-\left[ a_{TC}\epsilon^{abc}\tilde{U}_L^a\tilde{D}_L^b\tilde{\bar{U}}_R^c+a_U\left(\tilde{H}_1\tilde{D}_L^a - \tilde{H}_2\tilde{U}_L^a\right) \tilde{\bar{U}}^a_R+a_N\left(\tilde{H}_1\tilde{E}_L-\tilde{H}_2\tilde{N}_L\right) \tilde{\bar{N}}_R\right.\nonumber\\
&+&\left.a_E\left(\tilde{H}^{\prime}_1\tilde{E}_L-\tilde{H}^{\prime}_2\tilde{N}_L\right) \tilde{\bar{E}}_R+a_R \tilde{\bar{U}}_R^a \tilde{\bar{U}}_R^a \tilde{\bar{E}}_R + \frac{1}{2}M_D\bar{D}^a_R\bar{D}^a_R+c.c.\right]-M^2_Q\tilde{\bar{Q}}^a_L\tilde{Q}^a_L\nonumber\\
&-&M^2_U\tilde{\bar{U}}^a_R\tilde{U}^a_R - M^2_L\tilde{\bar{L}}_L\tilde{L}_L-M^2_N\tilde{\bar{N}}_R\tilde{N}_R-M^2_E\tilde{\bar{E}}_R\tilde{E}_R.
\ee

\end{document}